\documentclass[aps,prb,showpacs,twocolumn,superscriptaddress]{revtex4}

\usepackage{graphicx}
\usepackage{dcolumn}

\begin{document}


\title{The infrared Hall conductivity in optimally-doped
Bi$_2$Sr$_2$CaCu$_2$O$_{8+\delta}$}

\author{D. C. Schmadel}
\email[]{schmadel@physics.umd.edu}
\affiliation{Department of Physics, University of Maryland, College Park,
Maryland 20742 USA}

\author{G. S. Jenkins}
\email[]{schmadel@physics.umd.edu}
\affiliation{Department of Physics, University of Maryland, College Park,
Maryland 20742 USA}

\author{J. J. Tu}
\affiliation{Department of Physics, City University of New York, New York, New
York 11973 USA}
\affiliation{Department of Physics, Brookhaven National Laboratory, Upton, New
York 11973 USA}

\author{G. D. Gu}
\affiliation{Department of Physics, Brookhaven National Laboratory, Upton, New
York 11973 USA}

\author{Hiroshi Kontani}
\affiliation{Department of Physics, Nagoya University, Furo-cho, Nagoya 464-8602, Japan}

\author{H. D. Drew}
\affiliation{Department of Physics, University of Maryland, College Park,
Maryland 20742 USA}
\affiliation{Center of Superconductivity Research, University of Maryland,
College Park, Maryland 20742, USA}


\date{\today}

\begin{abstract}
The frequency dependence in the mid IR from 900 to 1100 cm$^{-1}$ and far IR at 84 cm$^{-1}$ and temperature dependence from 35 to 330 K of the normal state Hall transport is reported in single crystal, optimally doped Bi$_2$Sr$_2$CaCu$_2$O$_{8+\delta}$. The results show a nearly Drude behavior in the Hall conductivity $\sigma_{xy}$, which stands in contrast to the more complex extended Drude behavior for the longitudinal conductivity $\sigma_{xx}$. The mid IR Hall scattering rate increases linearly with temperature and has a small, positive, projected intercept of 200 cm$^{-1}$ at $T=0$. In contrast, the longitudinal scattering rate is much larger and exhibits very little temperature dependence. The spectral weight of $\sigma_{xy}$ is suppressed to 0.09 times the band value, whereas the spectral weight of $\sigma_{xx}$ is 0.33 times the band value. These disparate behaviors are consistent with calculations based on the fluctuation-exchange interaction when current vertex corrections are included.
\end{abstract}


\maketitle
While significant progress has been made toward an understanding of the cuprate superconductors due in large part to important new experimental data from ARPES, neutron scattering, and IR measurements, a clear microscopic theory remains elusive. Indeed, a theory of the superconductivity and transport properties of these strongly correlated materials continues as a central problem in modern condensed matter physics. The correlations with magnetic fluctuations have been elucidated by neutron scattering~\cite{SidisNeutrons} and the nature and occurrence of the quasiparticles near the Fermi surface and in different phases have been elucidated by ARPES~\cite{ShenARPES}. Evidence is accumulating that the underdoped cuprates, particularly the electron doped materials, support spin density waves which partially gap the Fermi surface~\cite{MillisZimmersSpinGap}. The optimally doped cuprates, however, are found to have a large nearly circular Fermi surface with quasiparticles that are well defined in the $(\pi,\pi)$ direction and somewhat less well defined in the $(\pi,0)$ direction. The linear temperature dependent resistivity is found to correspond to the linearly temperature dependent quasiparticle imaginary self-energy. At optimal doping the cuprates do not exhibit significant evidence for the pseudogap observed in the underdoped materials\cite{TimuskRPP62.61(1999)}.  Nevertheless, anomalous transport properties of the nearly optimally doped cuprates are universally observed.  One such anomaly is the temperature dependence of the Hall coefficient which has frequently been cited as evidence for non Fermi liquid character of the cuprates.    

One important new approach to understanding the magneto transport anomalies as well as other transport anomalies of the cuprates is to include vertex corrections in the conductivity within Fermi liquid theory.  Kontani and Yamada~\cite{KontaniVCwithConservation} have examined this approach using the spin fluctuation exchange interaction between carriers and include the current vertex corrections in the conductivity. They find that the current vertex corrections enter into $\sigma_{xy}$ more significantly in than $\sigma_{xx}$ and that this theory is capable of explaining the anomalous DC magneto-transport of the optimally doped cuprates in terms of the temperature dependence of the spin fluctuation induced interaction.  More recently, Kontani~\cite{KontaniVertexCorrections&Hall} has examined the IR frequency dependence of the magneto-transport with further success, which further highlights the possibility that the strong correlations may enter the diagonal and off-diagonal conductivities differently.  This insight suggests that $\sigma_{xy}$ may be a more useful quantity to study than the Hall angle, $\theta_H=\sigma_{xy}/\sigma_{xx}$. In this work we examine the temperature dependence of $\sigma_{xy}$ in optimally doped Bi$_2$Sr$_2$CaCu$_2$O$_{8+\delta}$ (BSCCO(2212)) which does not have conduction by chains, which complicated the interpretation of the IR Hall data for YBa$_2$Cu$_3$O$_7$ of earlier work~\cite{CernePRL2000}. We find that, for frequencies around 10 meV and 100 meV, $\sigma_{xy}$ is consistent with a nearly Drude form in contrast to both $\sigma_{xx}$ and $\theta_H$. The parameters of the Drude form are evaluated and compared with other transport properties, with theoretical calculations based on Kontani~\cite{KontaniVertexCorrections&Hall}, and with the $\sigma_{xy}$ sum rule. 

\begin{figure}
\includegraphics[width=8.6cm, clip=true]{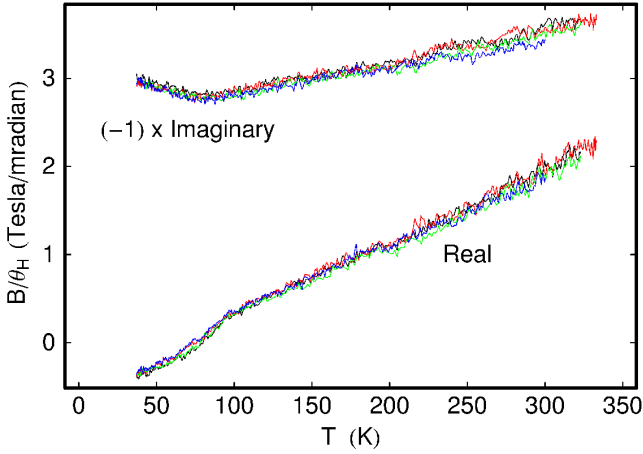}
\caption{\label{fig:dataInverseHallVsT} The real and imaginary parts of
the inverse Hall angle $\theta_{\text{H}}^{-1}$ for 2212 BSCCO versus temperature measured at 950 cm$^{-1}$ and normalized to one Tesla from four pairs of temperature scans.}
\end{figure}

The experimental system of the current work measures the very small complex
Faraday angle imparted to CO$_2$ laser radiation traveling perpendicular to and
transmitted by the sample immersed in a magnetic field. The system is the same as that used by in the earlier YBCO study~\cite{CernePRL2000} with the addition of an inline calibration system and a continuous stress-free temperature scan provision~\cite{Cerne2002}. The sample of the current work was cleaved, or rather peeled, from a bulk single
crystal of Bi$_2$Sr$_2$CaCu$_2$O$_{8+\delta}$ grown by the traveling
floating zone method~\cite{BSCCOFloatingZone}. The resulting 100 nm thick film was placed in thermal contact with a supporting wedged BaF$_2$ substrate.  Measurement of the AC magnetic susceptance of this mounted, peeled segment revealed a T$_{\text{c}}$ of 92 K with a width of less than 1K. This measurement was performed after all of the Hall measurements had been completed thus establishing the integrity of the sample and recommending the Hall data as representative of optimally doped BSCCO. The far infrared measurements were performed on a similarly peeled sample of BSCCO mounted onto a quartz substrate. The measurement apparatus used a molecular vapor laser as the source and a detection system similar to that of Grayson et al.\cite{GraysonPRL2002}Infrared conductivity data~\cite{TuPRB2002}, from measurements performed on bulk crystals from the same batch, supplied the real and imaginary parts of of the longitudinal conductivity $\sigma_{xx}$ required to obtain the Hall angle and the transverse conductivity $\sigma_{xy}$ from the measured Faraday angle~\cite{Cerne2002}. Fig.~\ref{fig:dataInverseHallVsT} shows the results for the mid infrared from 4 pairs of temperature scans.

We may analyze the results using the extended Drude model
 \begin{equation}
\label{eq:sigmaXX}
\sigma_{\text{xx}}=\frac{S_{xx}}{\gamma_{xx}-i\omega(1+\lambda(T,\omega))}=\frac{S^{*}_{xx}}{\gamma^{*}_{xx}-i\omega},
\end{equation}
\begin{equation}
\label{eq:sigmaXY}
\sigma_{\text{xy}}=\frac{S_{xy}}{(\gamma_{xy}-i\omega(1+\lambda(T,\omega)))^2}=\frac{S^{*}_{xy}}{(\gamma^{*}_{xy}-i\omega)^2}.
\end{equation}
where $S_{xx}$ and $S_{xy}$ are the longitudinal and transverse spectral weights. In this more generalized form the longitudinal and transverse scattering rates, $\gamma^{*}_{xx}$ and $\gamma^{*}_{xy}$, may be different. In fact, attempts to analyse the data under the assumption that they are identical leads to nonphysical results. Consider that with $\gamma^{*}_{xx}= \gamma^{*}_{xy}=\gamma^{*}$ then from $\theta_H=\sigma_{xy}/\sigma_{xx}$
\begin{equation}
\label{eq:thetaH}
\theta^{-1}_{H}=\frac{\gamma^{*}}{\omega^{*}_{H}}-\frac{\omega}{\omega^{*}_{H}},
\end{equation}
where $\omega^{*}_{H}=S^{*}_{xy}/S^{*}_{xx}$.
From Fig.~\ref{fig:dataInverseHallVsT} we see that Eq.~(\ref{eq:thetaH}) implies a normal state scattering rate with a negative projection below 50 K and a Hall mass of ~2.9 $m_{e}$, which, though similar in magnitude to the ARPES measured mass, increases with temperature. Such unconventional results challenge this simple analysis and suggest that the effective longitudinal and transverse scattering rates might be different. Therefore in this paper we examine the transverse conductivity $\sigma_{xy}$ independently.
\begin{figure}
\includegraphics[width=8.6cm, clip=true]{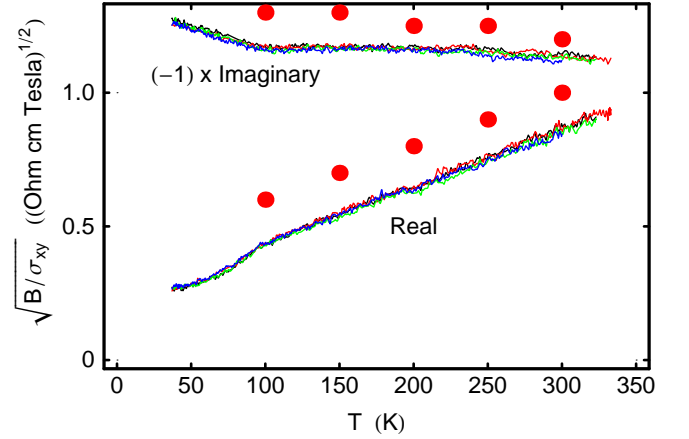}
\caption{\label{fig:inverseSquarerootSigmaVsT} $\sigma_{\text{xy}}^{-1/2}$  for 2212 BSCCO at 300 K versus temperature measured at 950 cm$^{-1}$ and normalized to one Tesla. The solid dots are the calculated results of the FLEX+CVC~\cite{KontaniFLEX+CVC} approximation}
\end{figure}

Fig.~\ref{fig:inverseSquarerootSigmaVsT} displays the data in terms of  $\sigma_{xy}$. The renormalized transverse scattering rate from Eq.~(\ref{eq:sigmaXY}) is simply
\begin{equation}
\label{eq:gammaXY}
\gamma_{\text{xy}}^*={-\omega}\frac{\text{Re}(\sigma_{\text{xy}}^{-1/2})}{\text{Im}(\sigma_{\text{xy}}^{-1/2})}.
\end{equation}
This alternative analysis of the data results in a transverse scattering rate $\gamma_{\text{xy}}^*$ which is everywhere positive, increases linearly with temperature, and has a zero temperature projection of ~200 cm$^{-1}$. This analysis also results in a transverse spectral weight which increase slightly with temperature. Remarkably, the observed temperature behavior of the transverse conductivity is well captured by the calculated results of the FLEX+CVC~\cite{KontaniFLEX+CVC} approximation for La$_{2-x}$Sr$_{x}$CuO$_{4}$ at 10 $\%$ doping shown as solid circles in Fig.~\ref{fig:inverseSquarerootSigmaVsT}. Similar results are obtained for YBa$_2$Cu$_3$O$_x$ calculations. In the relaxation time approximation, where the CVC is dropped, Re$(\sigma_{xy}^{-1/2}) \propto\gamma(\omega)$ and is approximately temperature independent for $\hbar\omega \gg k_BT$. According to the calculations, the approximate linear temperature behavior of Re$(\sigma_{xy}^{-1/2})$ is caused by the CVC due to spin fluctuations, which are strongly temperature dependent. Although the CVC is less effective as the doping increases, a similar numerical result is obtained even at 15\% doping if we use a larger value of the Coulomb interaction $U$, which is the single adjustable parameter of the theory. The existence of the CVC is a consequence of conservation laws, which intimately govern the transport phenomena. Consequently, neglecting the CVC frequently leads to nonphysical predictions. In these calculated results the vertex corrections enter into $\sigma_{xx}$ than $\sigma_{xy}$ quite differently.

\begin{figure}
\includegraphics[width=8.6cm, clip=true]{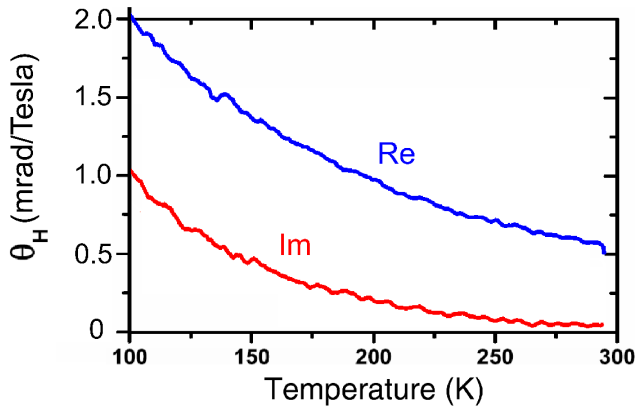}
\caption{\label{fig:ThetaHallAt84cmVsTcolor} $\theta_{\text{H}}$ for optimally doped 2212 BSCCO versus temperature measured at 84 cm$^{-1}$ averaged from 5 thermal scans and normalized to 1 Tesla.}
\end{figure}

\begin{figure}
\includegraphics[width=8.6cm, clip=true]{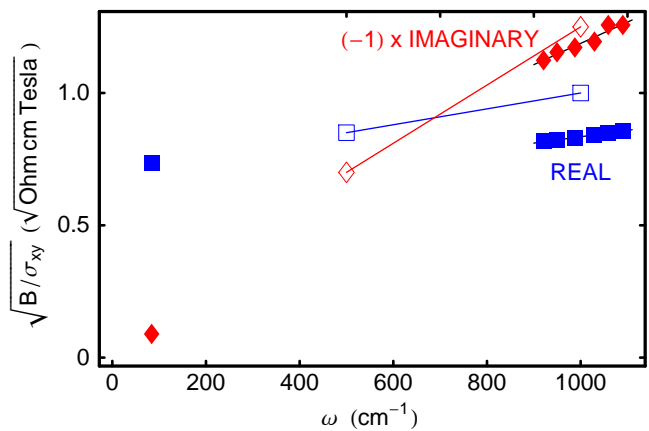}
\caption{\label{fig:inverseSquarerootSigmaVsF}$\sigma_{\text{xy}}^{-1/2}$  for 2212 BSCCO at 300 K and normalized to one Tesla. The data is in solid squares (real) and solid diamonds (imaginary) and the FLEX+CVC calculations are represented by open squares (real) and open diamonds (imaginary).}
\end{figure}

Equally remarkable, the FLEX+CVC approximation captures the frequency behavior of the transverse conductivity as well. To examine this frequency dependence we introduce the data shown in Fig.~\ref{fig:ThetaHallAt84cmVsTcolor} for optimally doped BSCCO at 84 cm$^{-1}$. Measurements were also performed at 24 and 42 cm$^{-1}$ and proved to be consistant with those shown. These far infrared data, multiplied by $\sigma_{xx}$ to produce $\sigma_{xy}$, appear in FIG.~\ref{fig:inverseSquarerootSigmaVsF} along with the mid infrared data and two frequency points from the FLEX+CVC calculations.

Since $\text{Im}(\sigma_{\text{xy}}^{-1/2})/\omega$ is found to be nearly independent of temperature and frequency at 950 cm$^{-1}$, it is interesting to consider the observed value in more general terms. The extended Drude model in Eq.~\ref{eq:sigmaXY}, relates $\text{Im}(\sigma_{\text{xy}}^{-1/2})$ to the Drude spectral weight $S_{\text{xy}}$~\cite{DrewColemanSumRule} defined here as:
\begin{equation}
\label{eq:Sxy}
{S_{\text{xy}}}=\int _{0}^{\Omega_c}{\frac{2}{\pi}} \omega\text{Im}\, \sigma_{\text{xy}}\, d\omega
\end{equation}
where $\Omega_c$ as a cut-off frequency high enough that the imaginary part of the quasiparticle self-energy is saturated but still below the Mott-Hubbard gap. The measured quantities in the IR Hall experiment at fixed frequency actually relate to 
$S_{\text{xy}}^{*}=(\omega/\text{Im}(\sigma_{\text{xy}}^{-1/2})^{2}=S_{\text{xy}}(1+\lambda_{xy})^{-2}$ and $\gamma_{\text{xy}}^{*}=\gamma_{\text{xy}}/(1+\lambda_{xy})$. Generally, for strongly interacting systems near a Mott transition, one expects $\lambda_{xy}\rightarrow0$ and $\gamma_{\text{xy}}$ to saturate at frequencies $\omega_s<\Omega_c$ as the measurement frequency exceeds the saturation frequency (typically $\sim$400 meV for cuprates) while still remaining below the Mott -Hubbard gap. This is the observed behavior of $\sigma_{xx}(\omega)$ in the optimally hole doped cuprates~\cite{vanderMarelNature425(2003)QuantumCritical}.

Assuming the same interaction energy scale for the transverse conductivity, the data allow us to charactierize the Drude peak in $\sigma_{xy}$, which is at a frequency $\omega\approx\gamma_{xy}^{*}$. We can compare $S_{xy}^{*}$, and $\gamma_{xy}^{*}$ at far IR and mid IR frequencies obtained from Eq.~\ref{eq:sigmaXY}. The results are shown in Table~\ref{tab:gammaSumTable}, which summarizes experimental results in terms of the scattering rates and spectral weights along with the longitudinal scattering rates from $\sigma^{*}_{xx}$. In the table, $\gamma_{\text{xy}}^*(84)$ is comparable to $\gamma_{\text{xx}}^*(84)$, however, $\gamma_{\text{xy}}^*(950)$ is much less than $\gamma_{\text{xx}}^*(950)$, corresponding to a weaker frequency dependence.
\begin{table*}
\caption{\label{tab:gammaSumTable}Renormalized scattering rates in cm$^{-1}$ for the longitudinal and tramsverse conductivity and the spectral weights in seconds$^{-3}$ per Tesla at far IR (84 cm$^{-1}$) and Mid IR (950 cm$^{-1}$) frequencies at different temperatures.}
\begin{ruledtabular}
\begin{tabular}{ccccccc}
T(K) & $\gamma_{\text{xx}}^*(84)$ \footnotemark[1]& $\gamma_{\text{xx}}^*(950)$ \footnotemark[1]& $\gamma_{\text{xy}}^*(84)$ & $\gamma_{\text{xy}}^*(950)$ & $S_{\text{xy}}^*(84)$ &  $S_{\text{xy}}^*(950)$\\
\hline
100 & 130 & 660 & 150 & 350 & $2.0 \times10^{40}$ & $2.1 \times10^{40}$ \\
200 & 400 & 920 & 400 & 530 & $2.4 \times10^{40}$ & $2.1 \times10^{40}$ \\
300 & 600 & 1070 & 690 & 720 & $2.8 \times10^{40}$ &  $2.3 \times10^{40}$\\
\end{tabular}
\end{ruledtabular}
\footnotetext[1]{Ref. ~\onlinecite{QuijadaBSCCO}.}
\end{table*}
Therefore, we expect $\lambda_{xy}<\lambda_{xx}$ and since $\lambda_{xx}<1$ at 950 $\text{cm}^{-1}$ we expect $\lambda_{xy}\ll1$. This suggests that $S_{\text{xy}}^{*}=S_{\text{xy}}(1+\lambda_{\text{xy}})^2\approx S_{\text{xy}}$ so that $S_{\text{xy}}^{*}$ in the mid IR should give a good approximate measure of the Drude contribution to the $\sigma_{xy}$ sum rule. Further support for this is obtained by comparing $S_{\text{xy}}^{*}$ in the mid and far IR from the table. It is seem that they are in good agreement again suggesting a very weak frequency dependence of the optical self energy $\Sigma_{xy}(\omega)=\omega\lambda_{xy}+ i\gamma_{xy}$ so that $\lambda_{xy}\ll1$. In fact these results indicate that $\sigma_{xy}$ has very nearly a simple Drude form in optimally doped BSCCO.

Since this analysis of the far IR and mid IR data appear to provide a measure of the partial sum $S_{\text{xy}}$ it is interesting to compare the value to the band value which can be calculated from the general relation~\cite{DrewColemanSumRule}, 
\begin{equation}
\label{eq:SxyBand}
{S_{\text{xy}}}=e^3B\sum_{k}\text{det}(m_{\text{k}}^{-1})n_k
\end{equation}
where
\begin{equation}
\label{eq:massTensor}
{m_k^{-1}}=\frac{\partial^2E(k)}{h^2 \partial k\partial k}
\end{equation}
is the inverse mass tensor and $n_k$ is the Fermi function. To calculate $S_{\text{xy}}^{band}$ the we use the tight binding fit to the cuprate band structure~\cite{MillisZimmersSpinGap}:
\begin{eqnarray}
\label{eq:MillisTightBindingFit}
E(k_x,k_y)=&&-2t_1(cos(k_x)+cos(k_y))\nonumber\\
&&+4t_2cos(k_x)cos(k_y)\nonumber\\
&&-2t_3(cos(2k_x)+cos(2k_y))
\end{eqnarray} 
where $t_1=0.38 \text{eV}$, $t_2=0.32t_1$, and $t_3=0.5t_2$. At optimal doping, the electron density $n$ is $0.84$ and $S_{\text{xy}}^{\text{band}}=3.73 \times10^{41} \text{Hz}^3/ \text{Tesla}$ and so $S_{\text{xy}}^{\text{*}}/S_{\text{xy}}^{\text{band}}=0.09$. It is interesting to compare this observed reduction from $S_{\text{xy}}$ with the behavior of $S_{\text{xx}}=\omega_{p}^{2}$/$4\pi$, where $\omega_p$ is the bare plasma frequency. Using the band model we find $S_{\text{xx}}^{\text{band}}=\frac{e^2}{2}\sum_{k}\text{tr}(m_{k}^{-1})n_k=3.50\times10^{30}{\text{ Hz}^2}$. From the literature~\cite{QuijadaBSCCO, TuPRB2002} $\omega_p=16200$ cm$^{-1}$ so that $S_{\text{xx}}^{\text{*}}/S_{\text{xx}}^{\text{band}}=0.33$. This compares with similar estimations for single layer Pr$_{2-x}$Ce$_{x}$CuO$_{4}$ and L$_{2-x}$Sr$_{x}$CO$_{4}$~\cite{MillisZimmersSpinGap}. Therefore we see that $S^{*}_{\text{xy}}$ is more significantly suppressed than $S^{*}_{\text{xx}}$. Indeed, $S_{\text{xy}}^{\text{*}}/S_{\text{xy}}^{\text{band}}\approx(S_{\text{xx}}^{\text{*}}/S_{\text{xx}}^{\text{band}})^2$. If interpreted as an effective mass then $S_{\text{xy}}/S_{\text{xx}}=\omega_{\text{H}}$ and $m_{\text{H}}/m_{\text{0}}=6.7$. The reduction in $S_{\text{xx}}$ is associated with the Coulomb correlations due to the proximity to the Mott transition as has been discussed recently~\cite{MillisZimmersSpinGap}. There are no theoretical results for $S_{\text{xy}}$. However, whether the reduction is to be thought of as a mass effect or a charge effect in Fermi liquid theory, a further reduction of $S_{\text{xy}}$ may be expected as $S_{\text{xx}}\sim e^2/m$ and $S_{\text{xy}}\sim e^3/m^2$. Therefore, theoretical prediction may be interesting and may give further insights in the strong correlations in the cuprates.

In summary, measurements of $\sigma_{\text{xy}}$ in the infrared exhibit a nearly Drude behavior with a scattering rate linear in temperature, and only weakly frequency dependent, and a nearly temperature and frequency independent transverse spectral weight, which is only 0.09 of the band value. These results are in good accord with calculations based on the fluctuation exchange model when current vertex corrections are included. Extending these IR Hall conductivity measurements over a wider frequency range in cuprates and other strongly correlated electron systems may provide significant new insights into the physics of these materials in the vicintity of a Mott transition.

We wish to thank A.J. Millis, S. Das Sarma, and V.M. Yakovenko for their stimulating discussions.  We also wish to thank D.B. Romero, who peeled and mounted the BSCCO sample. 

This work was supported by the NSF under grant DMR-0030112


\end{document}